# Bulk high-temperature superconductivity in the high-pressure tetragonal phase of bilayer La$_2$PrNi$_2$O$_7$

Ningning Wang[1,2#*], Gang Wang[1,2#], Xiaoling Shen[3,4#], Jun Hou[1,2], Jun Luo[1,2], Xiaoping Ma[1], Huaixin Yang[1,2], Lifen Shi[1,2], Jie Dou[1,2], Jie Feng[1,2], Jie Yang[1,2], Yunqing Shi[1,2], Zhian Ren[1,2], Hanming Ma[1,3], Pengtao Yang[1,2], Ziyi Liu[1,2], Yue Liu[1,2], Hua Zhang[1,2], Xiaoli Dong[1,2], Yuxin Wang[1,2], Kun Jiang[1,2], Jiangping Hu[1,2], Stuart Calder[5], Jiaqiang Yan[6], Jianping Sun[1,2], Bosen Wang[1,2], Rui Zhou[1,2*], Yoshiya Uwatoko[3*], and Jinguang Cheng[1,2*]

[1]*Beijing National Laboratory for Condensed Matter Physics and Institute of Physics, Chinese Academy of Sciences, Beijing 100190, China*

[2]*School of Physical Sciences, University of Chinese Academy of Sciences, Beijing 100190, China*

[3]*Institute for Solid State Physics, University of Tokyo, Kashiwa, Chiba 277-8581, Japan*

[4]*Key Laboratory of Artificial Structures and Quantum Control, School of Physics and Astronomy, Shanghai Jiao Tong University, Shanghai 200240, China*

[5]*Neutron Scattering Division, Oak Ridge National Laboratory, Oak Ridge, Tennessee 37831, USA*

[6]*Materials Science and Technology Division, Oak Ridge National Laboratory, Oak Ridge, Tennessee 37831, USA*

\# These authors contribute equally to this work.
*Corresponding authors: nnwang@iphy.ac.cn; rzhou@iphy.ac.cn; uwatoko@issp.u-tokyo.ac.jp; jgcheng@iphy.ac.cn

## Abstract

The Ruddlesden-Popper (R-P) bilayer nickelate, La$_3$Ni$_2$O$_7$, was recently found to show signatures of high-temperature superconductivity (HTSC) at pressures above 14 GPa[1]. Subsequent investigations achieved zero resistance in single- and poly-crystalline samples under hydrostatic pressure conditions[2-4]. Yet, obvious diamagnetic signals, the other hallmark of superconductors, are still lacking owing to the filamentary nature with low superconducting volume fraction[2,4,5]. The presence of a novel "1313" polymorph and competing R-P phases obscured proper identification of the phase for HTSC[6-9]. Thus, achieving bulk HTSC and identifying the phase at play are the most prominent tasks at present. Here, we address these issues in the praseodymium (Pr)-doped La$_2$PrNi$_2$O$_7$ polycrystalline samples. We find that the substitutions of Pr for La effectively inhibits the intergrowth of different R-P phases, resulting in nearly pure bilayer structure. For La$_2$PrNi$_2$O$_7$, pressure-induced orthorhombic-to-tetragonal structural transition takes place at $P_c \approx 11$ GPa, above which HTSC emerges gradually upon further compression. The superconducting transition temperatures at 18-20 GPa reach $T_c^{onset} = 82.5$ K and $T_c^{zero} = 60$ K, which are the highest values among known nickelate superconductors. More importantly, bulk HTSC was testified by detecting clear diamagnetic signals below ~75 K corresponding to an estimated superconducting volume fraction ~ 57(5)% at 20 GPa. Our results not only resolve the existing controversies but also illuminate directions for exploring bulk HTSC in the bilayer nickelates.

The recent report on the signature of HTSC in the pressurized $La_3Ni_2O_7$ crystals has reignited tremendous research interest in the R-P nickelates[1-30], $La_{n+1}Ni_nO_{3n+1}$ ($n$ = 1, 2, 3, ⋯, ∞), which consists of alternatively stacked $n$LaNiO$_3$ and LaO layers along $c$-axis. Since the Ni valence increases with $n$, stabilization of these R-P nickelates depends sensitively on the oxygen pressure ($p$O$_2$)[31]. Previous studies have demonstrated peculiar difficulties in obtaining single-phase $La_3Ni_2O_7$ crystals with a nominal $Ni^{2.5+}$ state[1,2,8,26,31,32]. It is only within a narrow range of $p$O$_2$ ≈ 10-18 bar that the "$La_3Ni_2O_7$" crystals with the majority $n$ = 2 (327) phase can be obtained[1,6,8,33], coexisting with some minority $n$ = 1 (214) and $n$ = 3 (4310) phases. For such $La_3Ni_2O_7$ crystals, a recent study reveals considerable oxygen vacancies at inner apical O1 positions[24]. In addition to the bilayer "2222" phase, recent structural analyses uncovered a novel "1313" polymorph featured by an alternative stacking of single- and triple-layer phases[6-9]; such "1313" crystals were found to show signatures of HTSC similar to the "2222" crystals[6]. These crystal imperfections and the presence of multiple structural variants obscured proper identification of the right phase responsible for HTSC. Under such circumstances, we recently resorted to the $La_3Ni_2O_{7-\delta}$ polycrystalline samples, which can be prepared in a relatively large quantity with better controlled quality and reproducibility via wet-chemistry sol-gel method[4,32]. This enabled us to perform comprehensive examinations on its crystal structure, oxygen content, and transport properties under HP. For this sample, we can reproducibly achieve zero resistance with $T_c^{zero}$ up to 40 K but did not observe discernable diamagnetic signals in ac magnetic susceptibility $\chi'(T)$ up to 15 GPa[4]. Our results reaffirmed the filamentary nature for the observed HTSC in $La_3Ni_2O_{7-\delta}$, while its origin remains an enigma. As shown below, these mysteries should root in the considerable structural disorders in $La_3Ni_2O_{7-\delta}$.

## I. Stacking faults in $La_3Ni_2O_{7-\delta}$ revealed by STEM and NQR

In our recent work, we noted that the ($h,k,0$) reflections of neutron powder diffraction (NPD) pattern of $La_3Ni_2O_{7-\delta}$ are resolution limited whereas those associated with $c$-axis display clear asymmetric Warren-like shapes, Fig. 1(a), which are typically attributed to short-range orders or stacking faults along the $c$-aixs[4]. To verify this, we performed HAADF-STEM on the $La_3Ni_2O_{7-\delta}$ samples. Although the bilayer structure dominates in most STEM images, our results revealed ubiquitous intergrowth of 327 with 4310 and 214 phases, as illustrated in Fig. 1(b), which is not surprising considering the similar structural arrangement of these R-P phases. How this intergrowth affects the transport properties of $La_3Ni_2O_{7-\delta}$, especially pressure-induced HTSC, is an intriguing issue that has been overlooked so far. To address this issue, we need to quantify the amount of stacking faults associated with the intergrowths. Apparently, this is challenging for local probes like STEM. Here, we employed nuclear quadrupole resonance (NQR) as a sensitive global probe to investigate the local structure disorders of bulk samples.

Due to the nuclear spin of $^{139}La$ being 7/2, we observed three sets of resonance peaks in the NQR spectrum (Extended Data Fig. 1). By comparing the quadrupole frequency $\nu_Q \approx$ 5-6 MHz with previous studies[25,34] (Extended Data Table 1), we determine that these three sets of resonance peaks correspond to the La(2) site. Notably, each set exhibits four peaks in the La(2) NQR spectrum, as illustrated in the upper panel of Fig. 1(c), indicating the presence of four distinct La(2) sites. By comparing to the NQR spectrum of $La_4Ni_3O_{10}$ shown in the middle panel of Fig. 1(c), the lowest frequency peak in the La(2) NQR spectrum of $La_3Ni_2O_{7-\delta}$ should correspond to that of $La_4Ni_3O_{10}$. In combination with the STEM results, we propose that the intergrowth between 327 and 4310 phases will result in two new sites: $La^{327i}(2)$ and $La^{4310i}(2)$, as depicted in Fig. 1(e). The $\nu_Q$ values for these two new sites are expected to fall between those of $La^{327}(2)$ and $La^{4310}(2)$, and their peaks should have equal areas owing to the same number of nuclei. Additionally, due to their positioning between 2- and 3-layer $NiO_2$ interfaces, the anisotropy parameter $\eta$ of the electrical field gradient (EFG) should become larger compared to $La^{327}(2)$ and $La^{4310}(2)$. As shown in Extended Data Table 1, the $\eta$ values for the middle two peaks are indeed much larger compared to those on either side. Moreover, the spectral weight ratio between $La^{327i}(2)$ and $La^{4310i}(2)$ is close to 1:1. These analyses indicate that the resonance peaks from low to high frequency in the upper panel of Fig. 1(c) correspond to $La^{4310}(2)$, $La^{4310i}(2)$, $La^{327i}(2)$, and $La^{327}(2)$ sites, respectively. We did not observe the 214 phase in our NQR spectra at 188 K because it is antiferromagnetically ordered[35,36], resulting in broad NQR lines and feeble signals [37]. Our study demonstrated that the $La_3Ni_2O_{7-\delta}$ polycrystalline samples contain a considerable amount of 327/4310 and 327/214 interfaces. Although the NQR spectrum reveals that the $La^{4310}(2)$ and $La^{4310i}(2)$ account for 30% of the NQR spectral weight in Fig. 1(c), the very small spectral weight of $La^{4310}(2)$ suggests that trilayer-$NiO_2$ does not exhibit continuous stacking along $c$-axis to form a real 4310 phase. This is elaborated by the STEM images in Fig. 1(b). Such short-range-ordered arrangement of 327/4310 and 327/214 interfaces along $c$-axis makes it hard to detect by conventional XRD. Whether a similar situation exists in the $La_3Ni_2O_7$ crystals deserves further scrutiny. This observation raises a critical question: are these stacking faults/interfaces responsible for the filamentary HTSC or detrimental for achieving bulk HTSC? To this end, it is essential to study 327-phase-pure samples. We show below that substitutions of $Pr^{3+}$ for $La^{3+}$ in $La_2PrNi_2O_{7-\delta}$ can effectively improve the phase purity.

## II. $La_2PrNi_2O_{7-\delta}$ with improved phase purity

A series of $La_{3-x}Pr_xNi_2O_{7-\delta}$ (0 ≤ x ≤ 1) polycrystalline samples were prepared via the sol-gel method (Details in Methods). Here we focus on $La_2PrNi_2O_{7-\delta}$ with the highest doping level that can maintain the orthorhombic *Amam* structure (Extended Data Fig. 2). In comparison to $La_3Ni_2O_{7-\delta}$, the asymmetric Warren-like feature of (0 0 10) peak is hardly discernable for $La_2PrNi_2O_{7-\delta}$, Fig. 1(a), signaling an obvious improvement of sample quality upon Pr doping. The refined unit-cell parameters of $La_2PrNi_2O_{7-\delta}$ are smaller than those of $La_3Ni_2O_{7-\delta}$ as expected, and the chemical pressure exerted by smaller $Pr^{3+}$ on the overall lattice corresponds to an external pressure of ~ 2.5 GPa according to the experimental *P*(*V*) relationship. Unlike physical pressure, the substitutions of $Pr^{3+}$ produce stronger structural distortions manifested by smaller Ni-O-Ni bond angles, which reduce the overlap integral between Ni-$e_g$ and O-2*p* orbitals and result in higher resistivity as observed experimentally (Extended Data Fig. 3). The EDX measurements confirm the uniform distribution of elemental compositions, while TGA reveals negligible oxygen deficiency δ ≈ 0.02(1), consistent with the NPD refinement results (Extended Data Fig. 4 and Table 2). Thus, we denoted the sample as $La_2PrNi_2O_7$ hereafter.

In accordance with the NPD results, the STEM images of $La_2PrNi_2O_7$ show that the 327/4310 intergrowth can hardly be observed, and the 327/214 intergrowth is significantly reduced, Fig. 1(d). The absence of the 4310 phase and corresponding 327/4310 interfaces was further verified by the NQR measurements. As shown in the lower panel of Fig. 1(c), the resonance peaks for $La^{4310}(2)$ and $La^{4310i}(2)$ are absent, while a broad peak is observed at the $La^{327}(2)$ position. The combined local STEM and global NQR results together with the NPD refinements unambiguously verified the improved phase purity with negligible oxygen vacancies in the $La_2PrNi_2O_7$ samples. In addition to the enhanced local structural distortions, we speculate that the valence instability of the Pr ions at elevated temperatures might be a key factor for the elimination of 4310 phase. More studies are underway to reveal the roles of Pr substitutions in $La_3Ni_2O_{7-\delta}$. The availability of nearly 327-phase-pure $La_2PrNi_2O_7$ samples provides an excellent opportunity to clarify the roles of the 327/4310 interfaces for pressure-induced HTSC mentioned above.

## III. Pressure-induced structural transition in $La_2PrNi_2O_7$

For $La_3Ni_2O_{7-\delta}$ crystal, the emergence of HTSC is concomitant with the orthorhombic *Amam* to *Fmmm* structure transition under HP[1]. Recent HP study reveals that the sample adopts a tetragonal *I4/mmm* structure in the superconducting phase at low temperatures[16]. It is interesting to examine whether structural transition appears in La$_2$PrNi$_2$O$_7$ under HP. Figure 2(a) shows the synchrotron XRD (SXRD) patterns collected at room temperature under various pressures from 3.2 to 56 GPa. The SXRD patterns below 10.3 GPa consistently match the orthorhombic *Amam* structure, as shown in Fig. 2(c) for the representative refinement at 3.2 GPa. However, upon compression to 11.1 GPa, several adjacent peaks are merged, such as the (020) and (200) peaks at ~13.4°, (135) and (315) peaks at ~23.1°, Fig. 2(b). This observation suggests pressure-induced structural transition towards a higher symmetry. Subsequent analyses revealed that the SXRD patterns of the HP phase can be better described using the Sr$_3$Ti$_2$O$_7$-type structural model with tetragonal *I4/mmm* space group (No. 139), Fig. 2(d), especially at pressure above 15 GPa (Extended Data Fig. 5).

As displayed in Fig. 2(e, f), the lattice parameters of La$_2$PrNi$_2$O$_7$ decrease smoothly with increasing pressure but exhibit anisotropic compressions. In the lower pressure range, the *b* decreases faster than *a* and they merge at 11.1 GPa, where the structural transition takes place. Such an orthorhombic-*Amam* to tetragonal-*I4/mmm* structural transition, Fig. 2(g), is distinct from that observed in La$_3$Ni$_2$O$_7$ crystals whose HP phase adopts the orthorhombic *Fmmm* symmetry at room temperature.

## IV. Emergence of HTSC in the HP tetragonal phase of La$_2$PrNi$_2$O$_7$

To explore potential HTSC in the HP tetragonal phase of La$_2$PrNi$_2$O$_7$, we measured resistivity $\rho(T)$ of sample #1 with cubic anvil cell (CAC) apparatus under various hydrostatic pressures up to 15 GPa. As shown in Fig. 3(a), $\rho(T)$ at 0 GPa exhibits a typical semiconducting behavior throughout the entire temperature range due to enhanced local structural distortions as discussed above. Upon compression, the magnitude of $\rho(T)$ decreases monotonically in the entire temperature range but retains a semiconducting-like behavior up to 5 GPa. When pressure is increased to 8 GPa, approaching the pressure of structural transition, a clear drop in $\rho(T)$ is observed below 21.4 K, Fig. 3(b). This behavior becomes more pronounced as pressure increases, signaling the appearance of superconductivity. At 12 GPa, $\rho(T)$ reaches zero resistance at 4.4 K with $T_c^{onset}$ = 66.4 K. $T_c^{zero}$ increases rapidly to 40 K at 15 GPa where $T_c^{onset}$ reaches 78.2 K. The emergence of HTSC in La$_2$PrNi$_2$O$_7$ is also coincident with the structural transition as observed in La$_3$Ni$_2$O$_7$[1].

To confirm reproducibility of the above results and to track $T_c$ to higher pressures, we measured $\rho(T)$ on sample #2 up to 20 GPa by employing two-stage 6/8 multianvil (MA) apparatus. As seen in Fig. 3(c, d), the obtained $\rho(T)$ at $P \leq 15$ GPa are consistent with those of sample #1, and the $\rho(T)$ at $P \geq 16$ GPa transform into metallic behavior with the normal state characterized by a $T$-linear strange-metal behavior over a wide temperature range. Similar behaviors were observed in La$_3$Ni$_2$O$_7$[1,3,4], implying an intimate relationship between the strange-metal behavior and HTSC in this system. Meanwhile, $T_c^{onset}$ increases to 82.5 K at 16 GPa and then decreases slightly to 80.4 K at 20 GPa, while $T_c^{zero}$ increases rapidly from 36.3 K at 15 GPa to ~60 K at 19-20 GPa. The negative effects of magnetic field and electrical current on $\rho(T)$ below $T_c^{onset}$ (Extended Data Fig. 6) further elaborate the occurrence of HTSC. Notably, La$_2$PrNi$_2$O$_7$ exhibits record-high $T_c^{zero} \approx 60$ K among known nickelate superconductors; it is about 20 K higher than that of La$_3$Ni$_2$O$_{7-\delta}$ at similar pressure. These results demonstrate that the absence of 327/4310 interfaces are beneficial for achieving a superior superconducting property, which is further substantiated by the observation of apparent diamagnetic signals below ~75 K.

## V. Evidence of bulk HTSC

Figure 3(e) displayed $\chi'(T)$ of La$_2$PrNi$_2$O$_7$ together with a piece of FeSe single crystal as reference. The diamagnetic response at the superconducting transition of FeSe at 2 and 9 GPa is clearly visible, in consistent with our previous studies[38,39]. Since FeSe undergoes a structural transition at about 10 GPa into a non-superconducting hexagonal phase[40,41], the observed diamagnetic signals in $\chi'(T)$ above 16 GPa should be attributed to La$_2$PrNi$_2$O$_7$. These diamagnetic responses can be better visualized by comparing with the background line superposed on each $\chi'(T)$ curve. As can be seen, the diamagnetic signal of La$_2$PrNi$_2$O$_7$ starts to appear below ~70 K at 16 GPa and gets stronger with increasing pressure. The superconducting shielding volume fraction, $f_{sc}$, of La$_2$PrNi$_2$O$_7$ can be estimated by comparing its diamagnetic response to that of FeSe at 9 GPa, which is confirmed to be $f_{sc} \approx 100\%$ (Extended Data Fig. 7). Considering the volume ratio of 3:2 between La$_2$PrNi$_2$O$_7$ and FeSe, the $f_{sc}$ of La$_2$PrNi$_2$O$_7$ at 8 K is estimated to increase rapidly from ~9.4(9)% at 16 GPa to ~57(5) % at 20 GPa.

We noted that the diamagnetic response in $\chi'(T)$ of La$_2$PrNi$_2$O$_7$ at $T_c$ is different from that of FeSe, which first drops sharply and then levels off. Such a distinct behavior indicates the gradual development of long-range coherence for Cooper pairing upon cooling down due to the polycrystalline nature of the studied La$_2$PrNi$_2$O$_7$ samples. The observation of $f_{sc} \sim 57(5)\%$ unambiguously testified to a bulk nature of observed HTSC in La$_2$PrNi$_2$O$_7$. Our work thus provides convincing evidence that the bulk HTSC originates from 327-phase-pure bilayer nickelates.

## VI. $T$-$P$ phase diagram

Figure 4 presents the constructed $T$-$P$ phase diagram of La$_2$PrNi$_2$O$_7$ sample, which depicts explicitly the emergence of HTSC in concomitant with the structural transition, similar to $La_3Ni_2O_{7-\delta}$[4] (Extended Data Fig. 8). Notably, bulk HTSC occurs at much higher pressures. Upon increasing pressure, $T_c^{onset}$ increases rapidly from 21.4 K at 8 GPa to 82.5 K at 16 GPa and then decreases slightly to 80.4 K at 20 GPa. The highest $T_c^{zero}$ reaches ~60 K at 19-20 GPa. The optimal $T_c^{onset}$ = 82.5 K and $T_c^{zero}$ = 60 K for $La_2PrNi_2O_7$ surpasses the corresponding values of $La_3Ni_2O_{7-\delta}$ at similar pressures. In addition, the slope of $dT_c^{zero}/dP$ ~ 10 K/GPa for $La_2PrNi_2O_7$ is much larger than that of ~ 4.5 K/GPa for $La_3Ni_2O_{7-\delta}$. These results indicate that the structural transition is a prerequisite to trigger superconductivity in the bilayer nickelates, while bulk HTSC requires higher pressure to further enhance the interlayer coupling between $3d_{z2}$ orbitals and to optimize the contribution of Ni-$3d_{z2}$ derived γ band near the Fermi level according to recent theoretical investigations[11,12,42]. The constructed T-P phase diagram also reveals a close relationship between strange-metal-like behavior and HTSC in the $La_2PrNi_2O_7$ samples.

## Conclusion

In summary, we demonstrated that the intergrowth of 327/4310 and 327/214 phases are detrimental to bulk HTSC in the bilayer $La_3Ni_2O_{7-\delta}$. By partially replacing La with Pr, we successfully inhibited the intergrowth of 327 with other R-P phases and achieved bulk HTSC with appreciable $f_{sc}$ ~ 57(5)% at 20 GPa in the HP tetragonal phase of $La_2PrNi_2O_7$. The optimal $T_c^{onset}$ = 82.5 K and $T_c^{zero}$ = 60 K are the highest values among nickelate superconductors reported so far. Our work indicates that $La_2PrNi_2O_7$ can be taken as an ideal platform to investigate the mechanism of HTSC for the bilayer nickelates.

## Methods

**Sample synthesis.** Polycrystalline $La_{3-x}Pr_xNi_2O_{7-\delta}$ (0 ≤ x ≤ 1) samples were synthesized with sol-gel method as described in previous studies[43,44]. Stoichiometric mixtures of $La_2O_3$, $Pr_6O_{11}$, and $Ni(NO_3)_2·6H_2O$, all of them with a purity of 99.99% from Alfa Aesar, were firstly dissolved in the deionized water with the addition of an appropriate amount of citric acid and nitric acid, and stirred in a 90 °C water bath for approximately 4 hours. Then the obtained vibrant green nitrate gel was heat treated overnight at 800°C to remove excess organic matter. After that, the product was ground, pressed to pellets, and sintered in air at 1100-1150 °C for 24 h.

**Sample characterizations.** The phase purity of $La_{3-x}Pr_xNi_2O_{7-\delta}$ at ambient condition was determined by XRD collected through Huber diffractometer with Cu $K_\alpha$ radiation. Temperature-dependent resistivity, $\rho(T)$, and magnetic susceptibility, $\chi(T)$, were measured by using Quantum Design Physical Properties Measurement System (PPMS) and Magnetic Property Measurement System (MPMS), respectively. Transmission electron microscope (STEM) and high-angle annular dark-field scanning transmission electron microscopy (HAADF-STEM) observations were performed in a JEOL ARM200F equipped with double-aberration correctors and operated at 200 kV. The STEM sample along [001] zone-axis direction was prepared by crushing the polycrystalline into fine fragments. Then the resulting suspensions were dispersed on holey copper grids. Thermogravimetric analysis (TGA) measurement was accomplished in NETZSCH STA 449F3, using a 10% $H_2$/Ar gas flow of 20 mL/min with a 5 K/min rate up to 1050 K. The $La_{3-x}Pr_xNi_2O_{7-\delta}$ samples tend to absorb moisture in air, we thus stored the samples in the glove box immediately after taking out from furnace. For TGA measurements, we carefully performed the blank tests. After TGA measurements, we measured XRD to confirm that the resultant products are $La_2O_3$, Ni, and $Pr_2O_3$. The analysis of chemical composition and microstructure utilized a Hitachi model S-4800 field emission scanning electron microscope (SEM) equipped with an energy-dispersive X-ray spectrometer (EDX). $^{139}$La-NQR spectra were measured on a powder sample weighing 300 mg by sweeping the frequency point by point and integrating spin-echo intensity.

Neutron powder diffraction (NPD) measurements were performed on the HB-2A diffractometer at the High Flux Isotope Reactor (HFIR), Oak Ridge National Laboratory (ORNL). Powder samples of $La_{3-x}Pr_xNi_2O_{7-\delta}$ were contained within a 6-mm-diameter vanadium sample can. Data were collected at 295 K with constant wavelength ($\lambda$ = 1.54 Å) measurements performed from the Ge (115) monochromator reflection. The NPD patterns were collected by scanning a 120° bank of 44 $^3$He detectors in 0.05° steps to give 2θ coverage from 5° to 150°. Rietveld refinements were performed with the FULLPROF program.

**High-pressure measurements.** The HP SXRD of $La_2PrNi_2O_7$ was measured at the 4W2 beamline at the Beijing Synchrotron Radiation Facility (BSRF) with a wavelength of $\lambda$ = 0.6199 Å. Rietveld analysis of HP SXRD data was performed with GSAS-II suite[45]. We employ the cubic anvil cell (CAC) and two-stage 6/8 multianvil (MA) apparatus to measure $\rho(T)$ at different pressures up to 15 and 20 GPa, respectively. Glycerol and Fluorinert FC70:FC77 (1:1) were employed as the liquid pressure transmitting medium (PTM) in CAC and MA, respectively. The pressure values inside the CAC and MA were estimated from pressure-loading force calibration curves determined by measuring the structure phase transitions of Bi, Sn, Pb, ZnS and GaAs at room temperature[46]. The ac magnetic susceptibility of $La_2PrNi_2O_7$ under various hydrostatic pressures up to 20 GPa was measured with the mutual induction method in the MA apparatus. The FeSe[38,39] single crystal was used as a reference for the superconducting diamagnetic signal because the lowest temperature of our MA apparatus can reach approximately 8 K, and the volume ratio of FeSe to $La_2PrNi_2O_7$ is about 2:3. An excitation currents of ~ 0.5 mA with a frequency of 732 Hz was applied to the excitation coil and the output signal was picked up with a Stanford Research SR830 lock-in amplifier.

## Data availability

All data are available from the corresponding authors upon reasonable request. Source data are provided with this paper.

## Acknowledgments

This work is supported by the National Key Research and Development Program of China (2023YFA1406100, 2021YFA1400200), National Natural Science Foundation of China (12025408, 11921004, U23A6003), the Strategic Priority Research Program of CAS (XDB33000000), the Specific Research Assistant Funding Program of CAS (E3VP011X61), the Postdoctoral Fellowship Program of China Postdoctoral Science Foundation (GZB20230828), the China Postdoctoral Science Foundation (2023M743740) and CAS PIFI program (2024PG0003). JQY was supported by the U.S. Department of Energy, Office of Science, Basic Energy Sciences, Division of Materials Sciences and Engineering. The high-pressure transport and the NQR experiments were respectively performed at the Cubic Anvil Cell station and the High Field Nuclear Magnetic Resonance station of Synergic Extreme Condition User Facility (SECUF). High-pressure synchrotron XRD measurements were performed at the 4W2 High Pressure Station, Beijing Synchrotron Radiation Facility (BSRF) and the BL15U1 station of Shanghai Synchrotron Radiation Facility (SSRF), which are supported by CAS. This research used resources at the High Flux Isotope Reactor, a U.S. DOE Office of Science User Facility operated by the Oak Ridge National Laboratory.

## Author contributions

J.G.C. designed and supervised this project. N.N.W. and G.W. synthesized the materials and characterized their structure via XRD and EDX; N.N.W., G.W., Y.L., H.Z. and X.L.D. measured the physical properties at ambient pressure; N.N.W., G.W. and J.H. performed high-pressure resistivity measurements by using the CAC apparatus with the support of H.M.M., P.T.Y., Z.Y. L., J.P.S. and B.S.W.; N.N.W., G.W. and L.F.S. measured the HP-SXRD; X.L.S. and U. W. performed the high-pressure resistivity and ac magnetic susceptibility measurements by using the MA apparatus; N.N.W., G.W. and J.G.C. analyzed all the collected data; J.L., J.D., J.F., J.Y. and R.Z. carried out the NQR measurements; R.Z. analyzed the NQR data; X.P.M., and H.X.Y. performed the

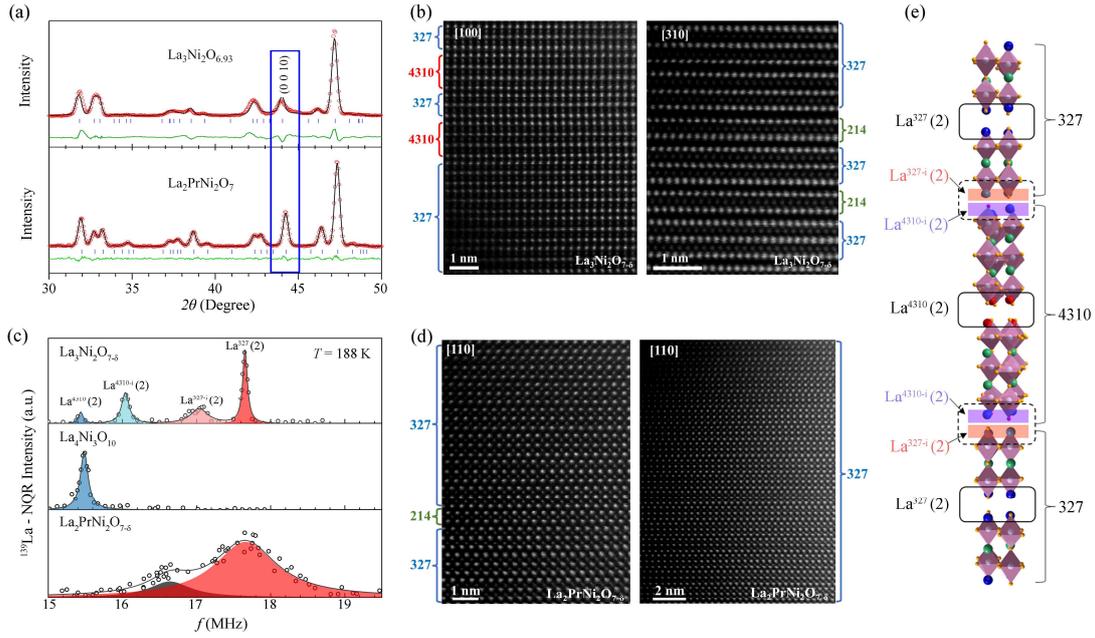

**Figure 1. Characterizations on the micro-structures of La$_{3-x}$Pr$_x$Ni$_2$O$_{7-\delta}$ (x = 0, 1) samples.** (a) Rietveld refinements of the NPD patterns in the 2θ range between 30° and 50°, highlighting the distinct features of the (0 0 10) reflection. The NPD data for x = 0 sample was taken from Ref. 4. High-angle annular dark field (HAADF) image taken along *c*-axis and the inset shows the selected area electron diffraction (SAED) pattern for (b) La$_3$Ni$_2$O$_{7-\delta}$ and (d) La$_2$PrNi$_2$O$_7$ sample. (c) $^{139}$La(2) NQR spectra around the frequency 3$\nu_Q$ in La$_3$Ni$_2$O$_{7-\delta}$, La$_4$Ni$_3$O$_{10}$ and La$_2$PrNi$_2$O$_7$ samples at 188 K. The solid lines represent fits obtained using Lorentz functions. The Pr doping inevitably induces structural disorders, resulting in a significant broadening of the NQR spectrum by an order of magnitude. No peaks corresponding to the 4310 phase are observed; only one broad peak associated with the 327 structure are present. A minor peak at 16.5 MHz might be attributed to the possible residual interfaces or the La sites adjacent to Pr. (e) A schematic drawing of the structure illustrating different La$^{4310}$(2), La$^{4310i}$(2), La$^{327i}$(2), and La$^{327}$(2) sites with triple-layer-NiO$_2$ intercalated in normal double-layer NiO$_2$ plane.

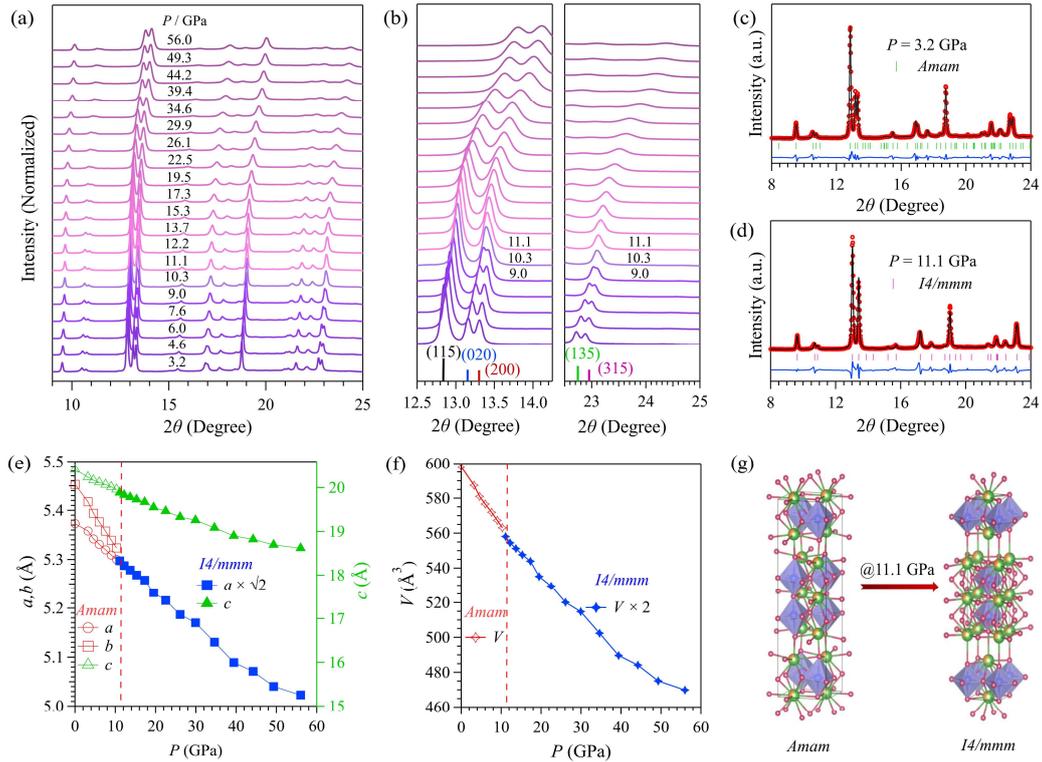

**Figure 2. Pressure-induced structural transition in La$_2$PrNi$_2$O$_7$.** (a) SXRD patterns of La$_2$PrNi$_2$O$_7$ powder samples under various pressures between 3.2 and 56 GPa. (b) The enlarged view of SXRD around the representative 2θ ranges, highlights the gradual merging of the diffraction peaks upon compression. (c) and (d) Refinement results of the SXRD patterns at 3.2 GPa using the space group *Amam* and at 11.1 GPa using the space group *I4/mmm*. (e) Lattice parameters and (f) Cell volume as a function of the pressure. (g) Crystal structure transformation of La$_2$PrNi$_2$O$_7$ under high pressure.

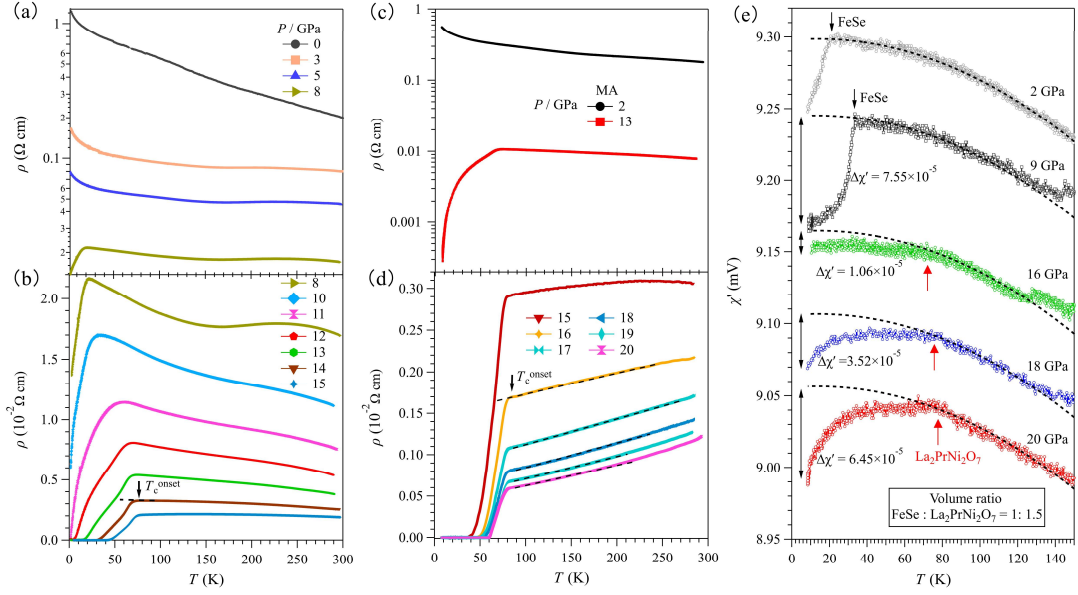

**Figure 3. Pressure-induced HTSC in the La$_2$PrNi$_2$O$_7$.** (a, b) $\rho(T)$ of the sample #1 under various hydrostatic pressures up to 15 GPa measured in CAC at the Institute of Physics CAS. Here, $T_c^{onset}$ is determined as the interception between two straight lines below and above the superconducting transition. (c, d) $\rho(T)$ of the sample #2 under various hydrostatic pressures up to 20 GPa measured in MA at the Institute for Solid State Physics (ISSP), University of Tokyo. (e) $\chi'(T)$ of sample #3 under various pressures up to 20 GPa measured with MA at ISSP. The data at pressure above 18 GPa shows clear diamagnetic signals below ~70-75 K, unambiguously confirming the achievement of bulk HTSC. The dashed line represents the background fitted from the data at 2 GPa.

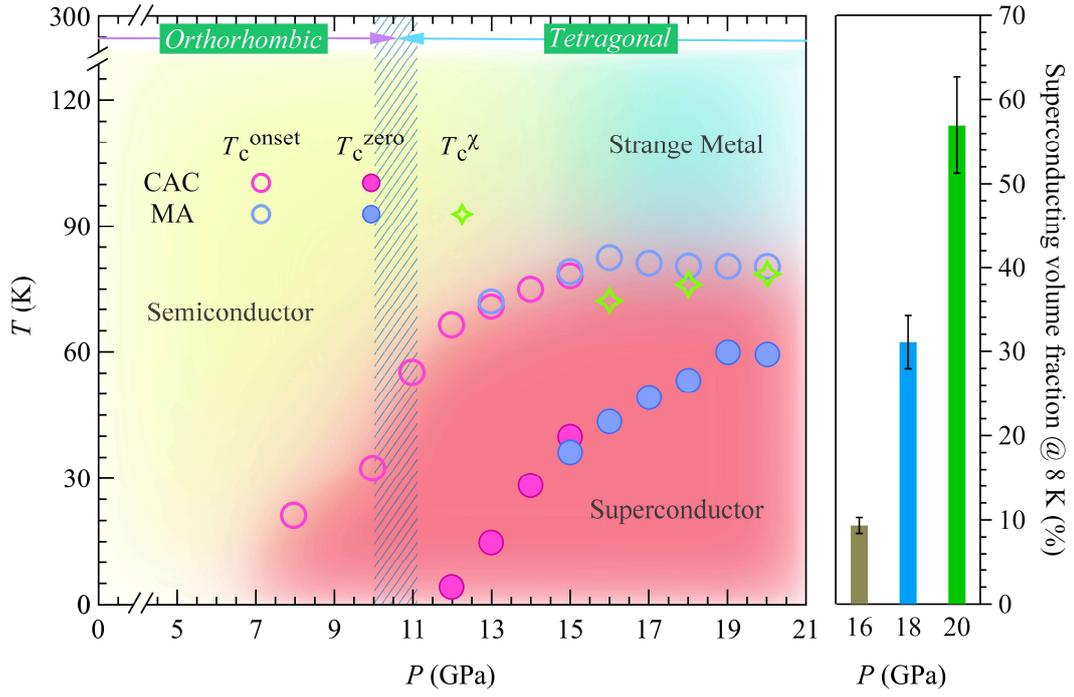

**Figure 4. The *T-P* phase diagram of La$_2$PrNi$_2$O$_7$.** The open and closed circles represent the onset and zero-resistance superconducting transition temperatures, $T_c^{onset}$ and $T_c^{zero}$, determined from the $\rho(T)$ measurements in CAC and MA. The star represents the $T_c^{\chi}$ determined from $\chi'(T)$ measurements in MA. The bar chart on the right represents the estimated superconducting shielding volume fraction, $f_{sc}$, at 8 K under different pressures.

**Extended Data Table 1. Experimental results of the quadrupole frequency $\nu_Q$, the asymmetry parameter $\eta$ of the La(2) sites at 188 K, and the deduced volume fraction from the Lorentz fitting in the upper panel of Fig. 1(c) for $La_3Ni_2O_{7-\sigma}$ polycrystalline sample.**

|  | $La^{327}(2)$ | $La^{327-i}(2)$ | $La^{4310-i}(2)$ | $La^{4310}(2)$ |
|---|---|---|---|---|
| $\nu_Q$ (MHz) | 5.886 | 5.676 | 5.359 | 74 |
| $\eta$ | 0.043 | 0.105 | 0.155 | 0.085 |
| Fraction (%) | 40.4±4.1 | 27.0±3.0 | 26.0±2.8 | 6.6±1.0 |

**Extended Data Table 2. NPD refinement results of La$_{3-x}$Pr$_x$Ni$_2$O$_{7-\delta}$ (x = 0.3, 1).** Crystal structure parameters for La$_{3-x}$Pr$_x$Ni$_2$O$_7$ (x = 0.3, 1) from NPD data at 295 K. All these values were extracted from the results of Rietveld refinements performed by using the FULLPROF program, where the atom occupancy was defined as the site multiplicity of each Wyckoff position divided by the maxima multiplicity of the space group. The maxima multiplicity of the space group *Amam* is 16.

**La$_{2.7}$Pr$_{0.3}$Ni$_2$O$_{7-\delta}$**

| Atom | Wyck | x | y | z | 100$U_{iso}$(Å$^2$) | Occupancy |
|---|---|---|---|---|---|---|
| La1 | 4c | 0.25 | 0.2503(21) | 0.5 | 0.093(82) | 0.225 |
| La2 | 8g | 0.25 | 0.2495(13) | 0.3171(1) | 0.513(58) | 0.45 |
| Pr1 | 4c | 0.25 | 0.2503(21) | 0.5 | 0.093(82) | 0.025 |
| Pr2 | 8g | 0.25 | 0.2495(13) | 0.3171(1) | 0.513(58) | 0.05 |
| Ni | 8g | 0.25 | 0.2531(14) | 0.0954(1) | 0.490(35) | 0.5 |
| O1 | 4c | 0.25 | 0.2964(18) | 0 | 0.973(32) | 0.231(3) |
| O2 | 8g | 0.25 | 0.2082(10) | 0.2017(3) | 0.973(32) | 0.5 |
| O3 | 8e | 0 | 0.5 | 0.1048(3) | 0.973(32) | 0.5 |
| O4 | 8e | 0.5 | 0 | 0.0875(2) | 0.973(32) | 0.5 |

Lattice parameters: $a$ = 5.39968(12) Å, $b$ = 5.45906(13) Å, $c$ = 20.50879(49) Å
$R_p$ = 5.31%, $R_{wp}$ = 6.63%, $\chi^2$ = 1.76, $R_{Bragg}$ = 4.42

**La$_2$PrNi$_2$O$_7$**

| Atom | Wyck | x | y | z | 100$U_{iso}$(Å$^2$) | Occupancy |
|---|---|---|---|---|---|---|
| La1 | 4c | 0.25 | 0.2483(13) | 0.5 | 0.300(68) | 0.167 |
| La2 | 8g | 0.25 | 0.2583(8) | 0.3183(1) | 0.615(44) | 0.333 |
| Pr1 | 4c | 0.25 | 0.2483(13) | 0.5 | 0.300(68) | 0.083 |
| Pr2 | 8g | 0.25 | 0.2583(8) | 0.3183(1) | 0.615(44) | 0.167 |
| Ni | 8g | 0.25 | 0.2513(8) | 0.0961(1) | 0.446(25) | 0.5 |
| O1 | 4c | 0.25 | 0.3013(11) | 0 | 0.757(23) | 0.242(2) |
| O2 | 8g | 0.25 | 0.2076(7) | 0.2046(2) | 0.757(23) | 0.5 |
| O3 | 8e | 0 | 0.5 | 0.1063(2) | 0.757(23) | 0.5 |
| O4 | 8e | 0.5 | 0 | 0.0876(2) | 0.757(23) | 0.5 |

Lattice parameters: $a$ = 5.38206(9) Å, $b$ = 5.46208(9) Å, $c$ = 20.43435(33) Å
$R_p$ = 4.67%, $R_{wp}$ = 5.63%, $\chi^2$ = 1.25, $R_{Bragg}$ = 2.49

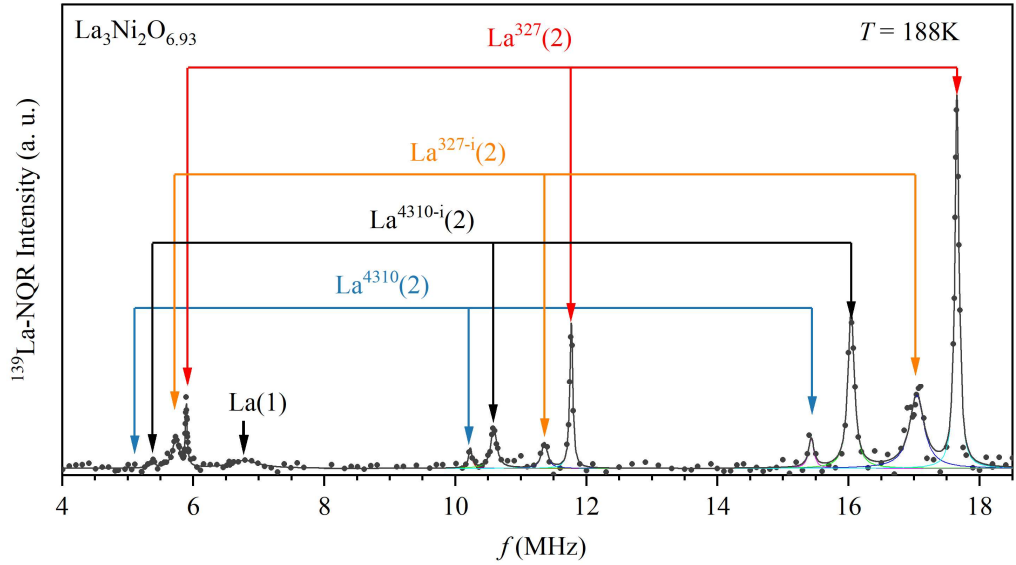

**Extended Data Fig. 1 $^{139}$La-NQR spectrum for the La$_3$Ni$_2$O$_{7-\delta}$ polycrystalline sample at 188 K.** Four distinct pairs of resonance peaks are observed, denoted by four sets of arrows, indicating the existence of four unique La(2) sites within the sample. Due to a broader linewidth, only one resonance peak corresponding to 7/2 – 5/2 transition is observed in the spectrum for the La(1) site. The solid lines represent fits obtained using Lorentz functions.

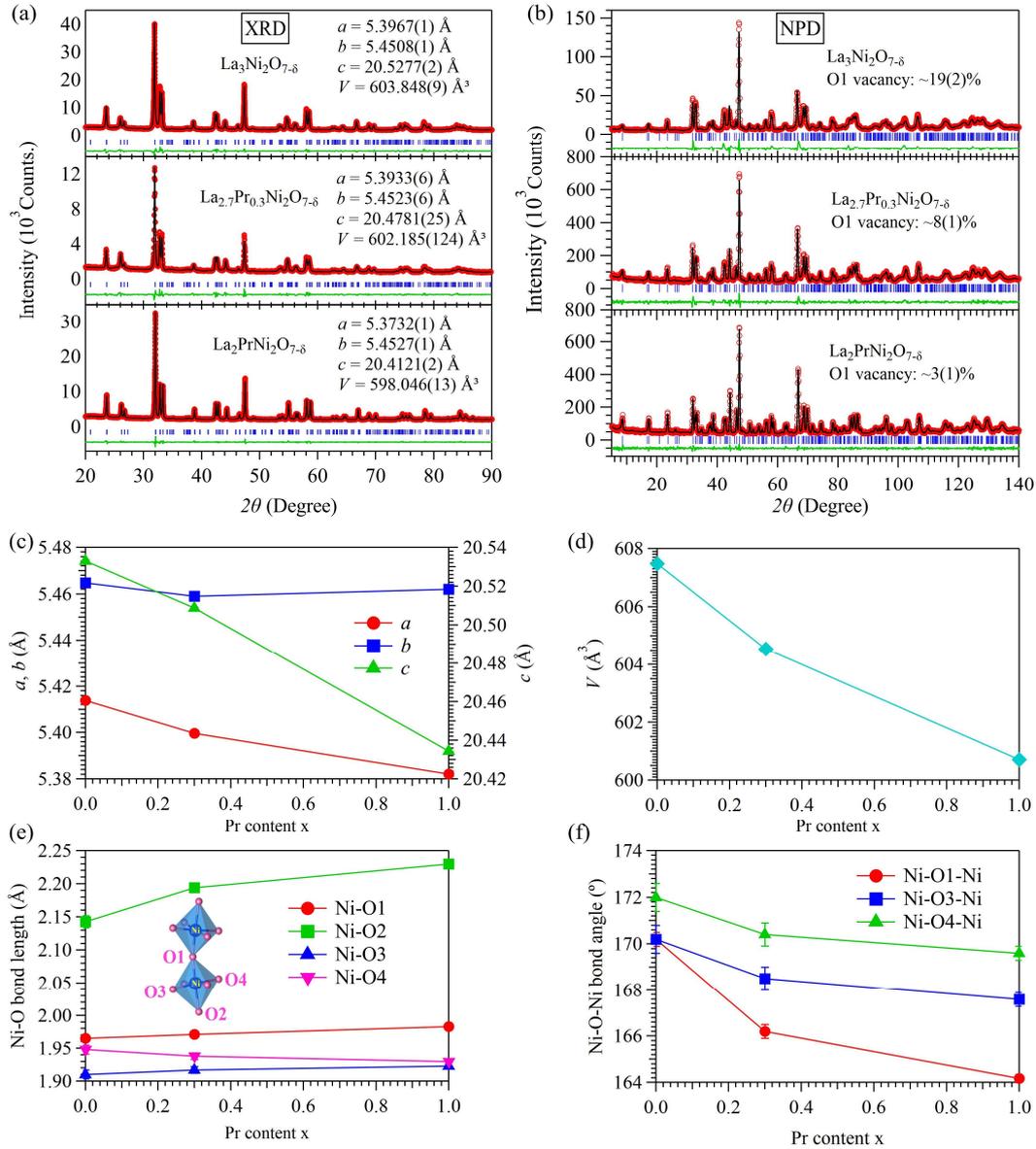

**Extended Data Fig. 2 Rietveld refinements on the XRD and NPD data of La$_{3-x}$Pr$_x$Ni$_2$O$_{7-\delta}$ (x = 0, 0.3, 1.0) polycrystalline samples.** Refinement results with the space group *Amam* of (a) XRD and (b) NPD patterns. (c) The obtained unit-cell parameters, and (d) volume as a function of the Pr-content x from NPD data. (e) Ni-O-Ni bond lengths and (f) Ni-O-Ni bond angles as a function of the Pr-content x from NPD data. According to the refinement results of NPD data, no site preference for the Pr was detected and the oxygen vacancies at the inner apical O1 sites decrease gradually with increasing the Pr content.

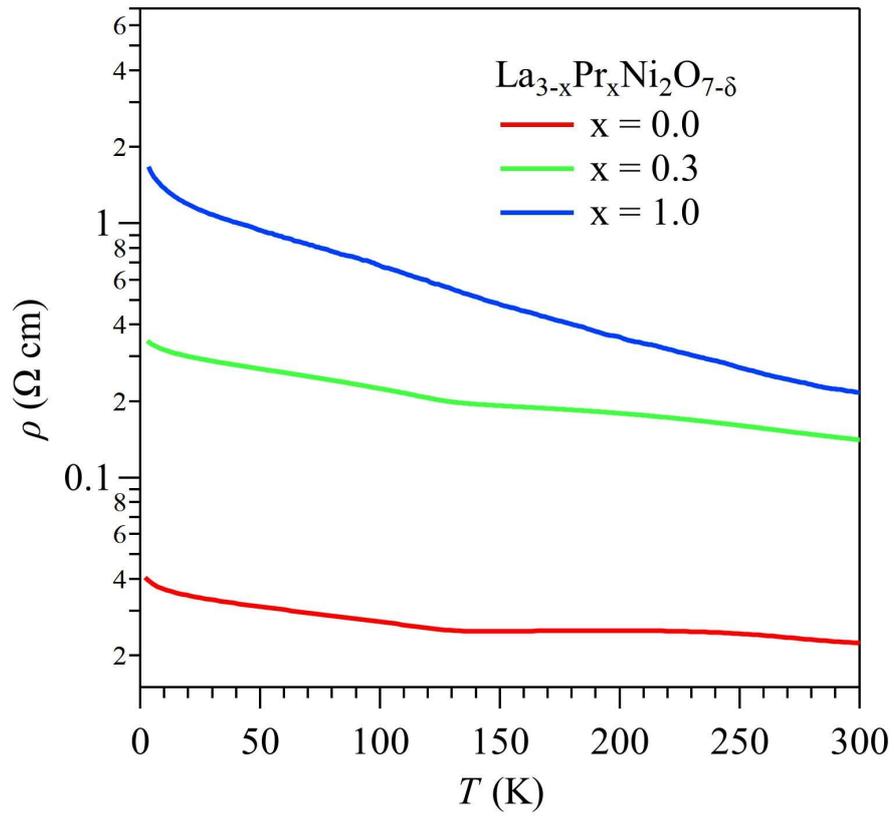

**Extended Data Fig. 3 Temperature dependence of resistivity $\rho(T)$ of La$_{3-x}$Pr$_x$Ni$_2$O$_{7-\delta}$ (x = 0, 0.3, 1.0) samples at ambient pressure.** The $\rho(T)$ increases gradually with increasing the Pr-content x.

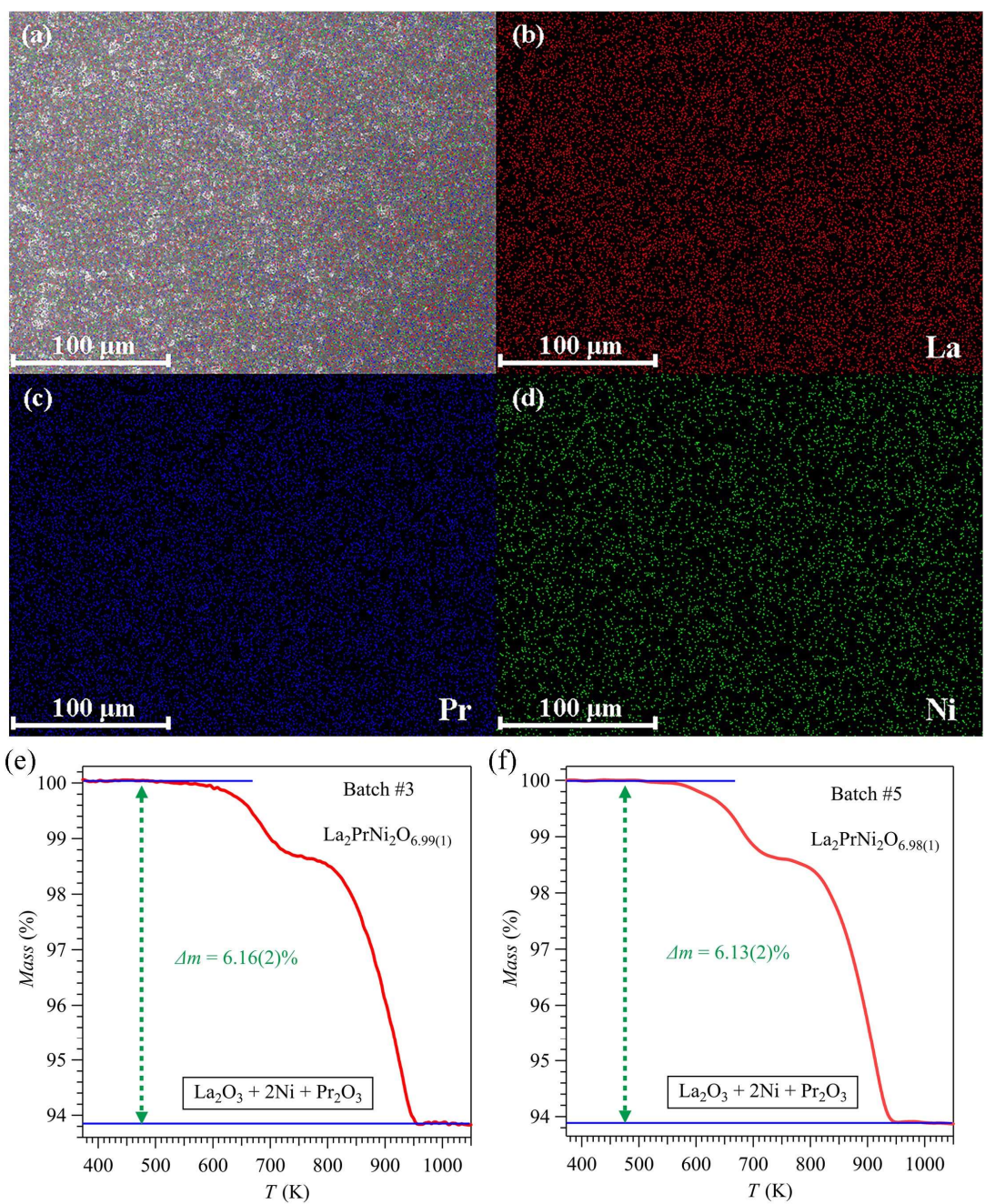

**Extended Data Fig. 4 EDX and TGA results of the La$_2$PrNi$_2$O$_7$ sample.** (a)-(d) Energy-dispersive X-ray spectroscopy (EDX) mapping patterns of La$_2$PrNi$_2$O$_7$ sample showing the uniform distribution of La, Pr, Ni elements. Each color represents a specific element, the mapping image illustrates the spatial distribution of elements within the sample. (e) and (f) The thermal decomposition behavior is similar to that in La$_3$Ni$_2$O$_{7-\delta}$ in Ref [4]. Analysis of the TGA data revealed a negligible oxygen deficiency with $\delta \approx 0.02(1)$ for La$_2$PrNi$_2$O$_{7-\delta}$, which is smaller than the $\delta \approx 0.07$ for La$_3$Ni$_2$O$_{7-\delta}$ prepared in the similar conditions.

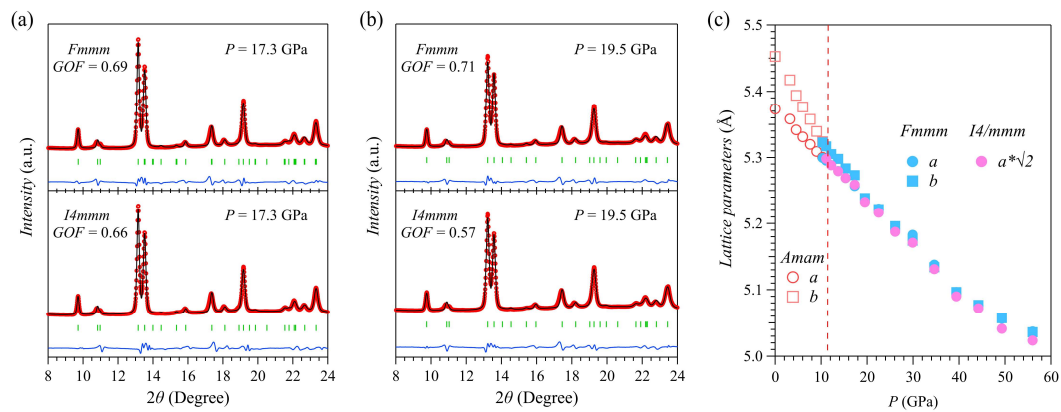

**Extended Data Fig. 5 GSAS-II refinement results on HP-SXRD of La$_2$PrNi$_2$O$_7$ samples by using orthorhombic *Fmmm* and tetragonal *I4/mmm* space group.** Rietveld refinements on the HP-SXRD patterns of La$_2$PrNi$_2$O$_7$ by using orthorhombic *Fmmm* and tetragonal *I4/mmm* space group under (a) 17.3 GPa and (b) 19.5 GPa. (c) Comparison of the obtained lattice parameters (*a* and *b*) as a function of the pressure. Although both space groups can refine the data equally well, the refinement results using the orthorhombic *Fmmm* space group at higher pressures show that *a* and *b* merge together, indicating the symmetry of the crystal structure has changed.

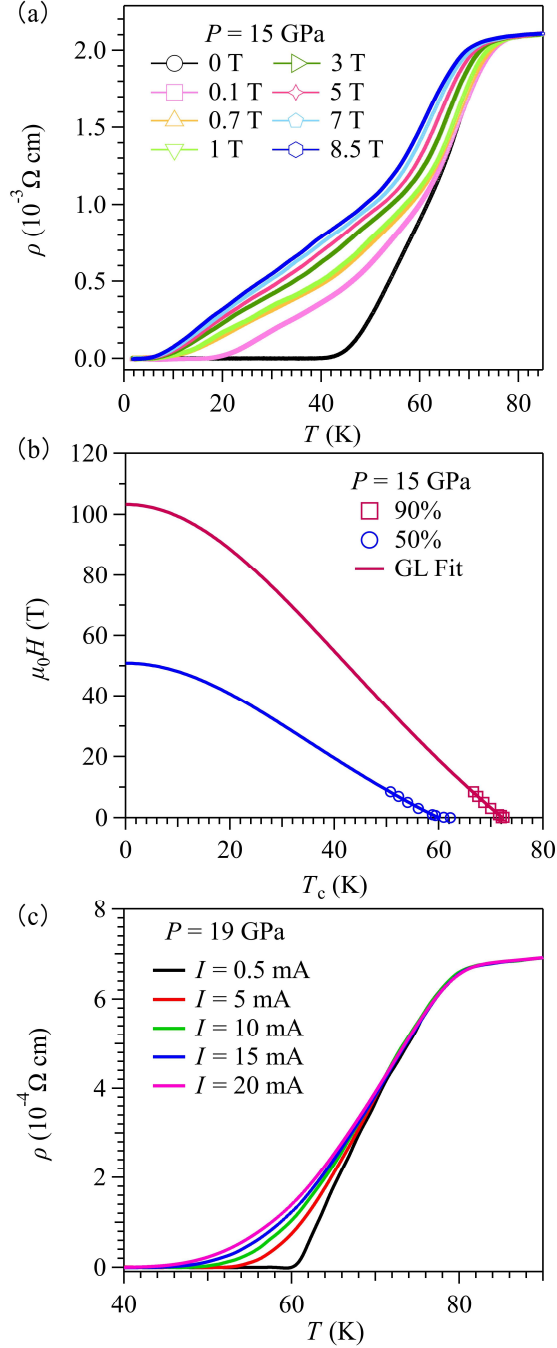

**Extended Data Fig. 6 The effects of magnetic field and electrical current on the superconducting transition of La$_2$PrNi$_2$O$_7$ under high pressures.** (a) The low-temperature resistivity $\rho(T)$ at 15 GPa under various magnetic fields up to 8.5 T. (b) Temperature dependence of the upper critical field $\mu_0 H_{c2}(T)$ at 15 GPa. The solid line is the fitting curve by using the formula $H_{c2} = H_{c2}(0)(1-t^2)/(1+t^2)$, where $t = T/T_c$. (c) The low-temperature resistivity $\rho(T)$ at 19 GPa of La$_2$PrNi$_2$O$_7$ polycrystalline sample #2 measured with different currents, which shows that $T_c^{zero}$ can be gradually inhibited by increasing electrical currents.

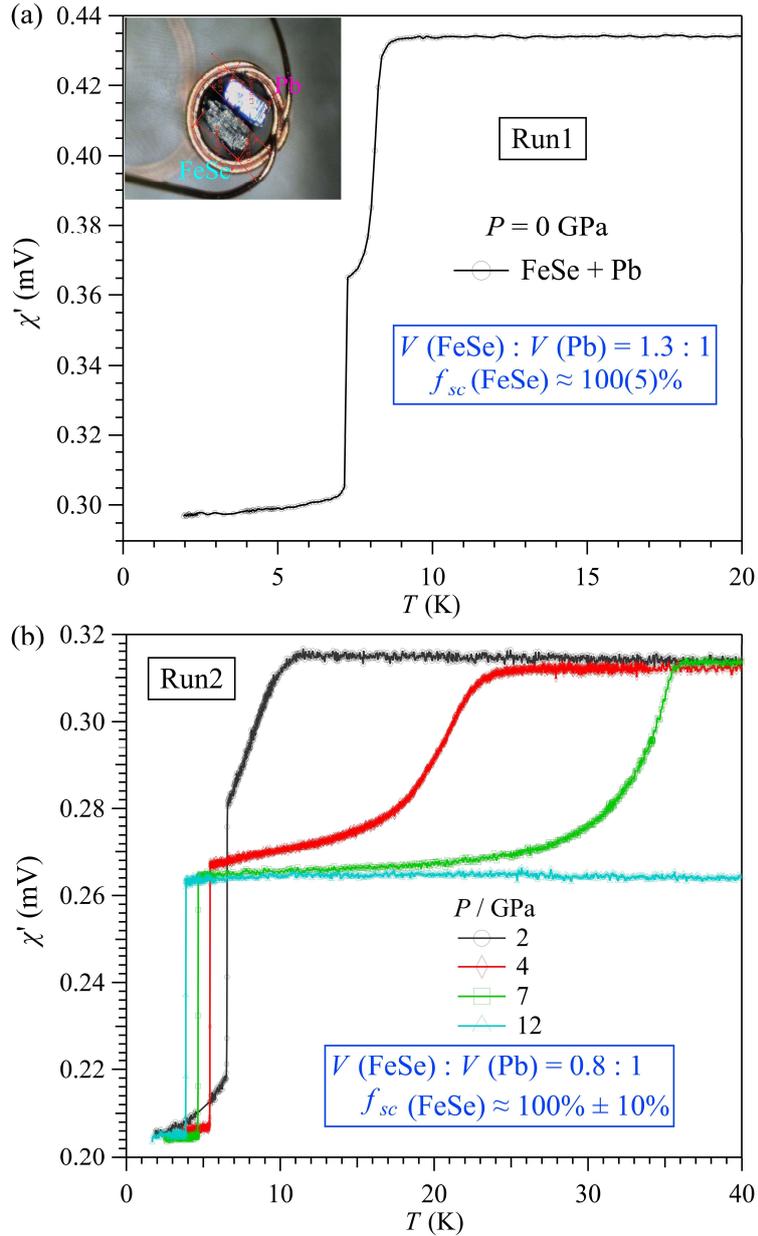

**Extended Data Fig. 7 Temperature dependence of ac magnetic susceptibility $\chi'(T)$ of FeSe single crystal.** (a) The $\chi'(T)$ data of the FeSe single crystals together with a piece of Pb measured at ambient pressure (Run1) and (b) under hydrostatic pressures up to 12 GP with the mutual induction method in CAC (Run2). The inset of (a) shows the photo of the pick-up coil filled with FeSe and Pb for Run1.

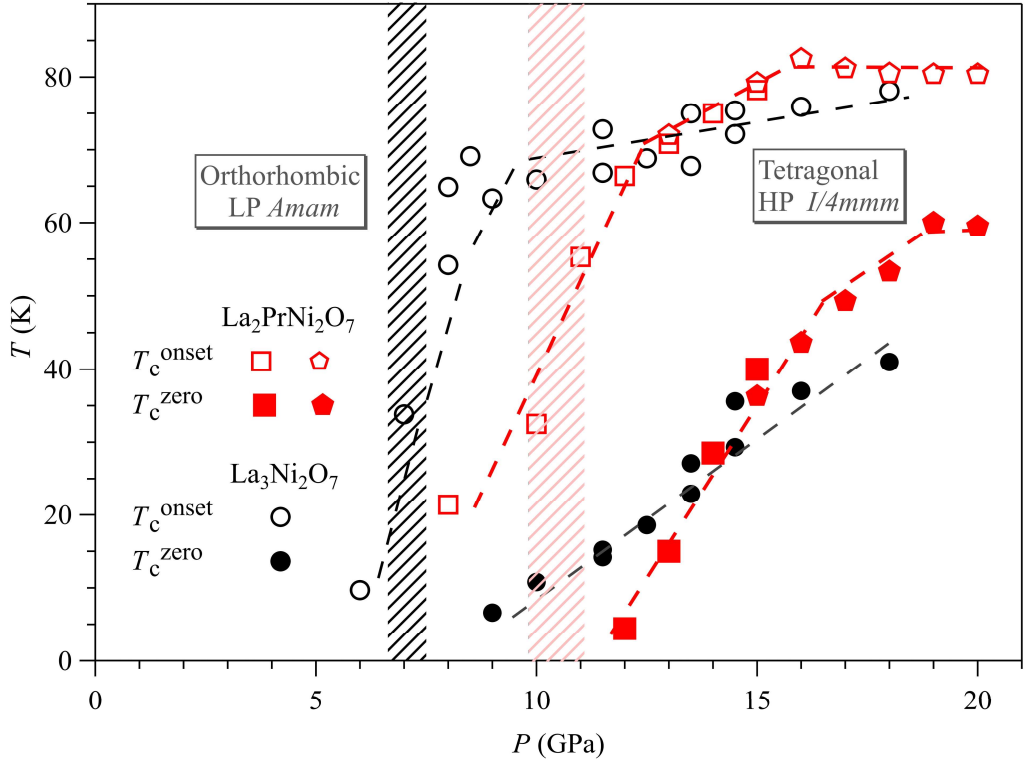

**Extended Data Fig. 8 Pressure dependence of $T_c$ for the La$_2$PrNi$_2$O$_7$ in comparison with that of La$_3$Ni$_2$O$_{7-\delta}$ polycrystalline samples.** The filled squares and pentagons represent the onset and zero-resistance superconducting transition temperatures of La$_2$PrNi$_2$O$_7$ determined from the $\rho(T)$ measurements in CAC and MA. The open marks for La$_3$Ni$_2$O$_{7-\delta}$ are taken from our previous study[4].